\documentclass[conference]{IEEEtran}
\IEEEoverridecommandlockouts
\usepackage{cite}
\usepackage{amsmath,amssymb,amsfonts}
\usepackage{algorithmic}
\usepackage{graphics} 
\usepackage{epsfig} 
\usepackage[abs]{overpic}
\usepackage{algorithmic}
\usepackage{epstopdf}
\usepackage{caption}
\usepackage{subcaption}
\graphicspath{{figs/}}
\usepackage{tikz}
\usetikzlibrary{arrows}
\usepackage[utf8]{inputenc}
\usepackage{mathabx}
\usepackage{titlesec}
\usepackage{textcomp}
\usepackage{xcolor}
\usepackage{tabto}
\def\BibTeX{{\rm B\kern-.05em{\sc i\kern-.025em b}\kern-.08em
    T\kern-.1667em\lower.7ex\hbox{E}\kern-.125emX}}
\begin{document}

\title{Emergence and Stability of Self-Evolved Cooperative Strategies using Stochastic Machines}

\author{\IEEEauthorblockN{Jin Hong Kuan}
\IEEEauthorblockA{\textit{Department of Computer Science} \\
\textit{University of Minnesota}\\
kuan0019@umn.edu}
\and
\IEEEauthorblockN{Aadesh Salecha}
\IEEEauthorblockA{\textit{Department of Computer Science} \\
\textit{University of Minnesota}\\
salec006@umn.edu}
}

\IEEEoverridecommandlockouts
\IEEEpubid{\makebox[\columnwidth] {978-1-7281-2547-3/20/\$31.00~\copyright2020 IEEE \hfill} \hspace{\columnsep}\makebox[\columnwidth]{ }}

\titlespacing*{\section}{0pt}{0.5ex  minus .2ex}{0.5ex  minus .2ex}
\titlespacing*{\subsection}{0pt}{0.5ex minus .2ex}{0.5ex minus .2ex}
\setlength{\textfloatsep}{0.1cm}

\maketitle

\IEEEpubidadjcol

\begin{abstract}
To investigate the origin of cooperative behaviors, we developed an evolutionary model of sequential strategies and tested our model with computer simulations. The sequential strategies represented by stochastic machines were evaluated through games of Iterated Prisoner's Dilemma (IPD) with other agents in the population, allowing co-evolution to occur. We expanded upon past works by proposing a novel mechanism to mutate stochastic Moore machines that enables a richer class of machines to be evolved. These machines were then subjected to various selection mechanisms and the resulting evolved strategies were analyzed. We found that cooperation can indeed emerge spontaneously in evolving populations playing iterated PD, specifically in the form of trigger strategies. In addition, we found that the resulting populations converged to evolutionarily stable states and were resilient towards mutation. In order to test the generalizability of our proposed mutation mechanism and simulation approach, we also evolved the machines to play other games such as Chicken, Stag Hunt, and Battle, and obtained strategies that perform as well as mixed strategies in Nash Equilibrium.
\end{abstract}


\section{Introduction}


There are many paradoxes in the natural world; among them is the curious case of the emergence of cooperation in an unforgiving Darwinian world. The emergence of cooperation has been addressed in various forms across different disciplines. John Maynard Smith developed the concept of evolutionarily stable strategies (ESS) to explain the instability of selfish behaviors \cite{smith_evolution_1982}. George Price vindicated the mathematical origins of altruism by showing its contributions to gene propagation \cite{price_extension_1972}. Richard Dawkins in \emph{The Selfish Gene} built upon earlier works in asserting that natural selection is oriented around the survival of the gene, rather than the individual, describing kin selection in terms of the gene's self-interest \cite{dawkins_selfish_2006}. 

Game theorists, on the other hand, have studied cooperation in more abstract and generalized terms. The tension between selfishness and optimality was explored in the form of the Prisoner's Dilemma (PD) by Merrill Flood, in which the payoffs were assigned to make the non-cooperative Nash equilibrium dominate the Pareto-superior cooperative solution \cite{flood1958some}. In such a scenario, the \emph{optimal} solution requires that all agents behave \emph{irrationally} and against their own interests. This conundrum was found to be resolvable when repeated play was introduced into the equation, as the Folk Theorem proved that any feasible and rationalizable strategy, including cooperation, can be a Nash Equilibrium strategy in an infinitely or indefinitely long game \cite{dutta1995folk}. This begs the question: Does there exist an iterated strategy that works well when pitted against any other strategy in a finite setting? One attempt to answer this question was carried out by Robert Axelrod in his famous tournaments, in which he gathered sequential strategies from 200 game theorists to play the Iterated Prisoner's Dilemma (IPD) and determine what strategy works best in relation to other strategies, and in a sense, is the most adaptable \cite{axelrod_evolution_1984}. He found that the cooperative strategy of Tit-for-Tat was able to outperform a variety of more complex strategies. 

In this paper, we look to evolution, specifically Darwinian natural selection to help us develop such strategies, while providing hypotheses about the rationale behind cooperative behaviors. 
We model iterated strategies as evolving automata, a framework pioneered by Fogel in the 1960s \cite{fogel_artificial_1966}. His work showed that by defining an evaluation objective and developing a set of mutation and survival mechanisms, one can induce the evolution of automata with specific behaviors. When applied to the study of IPDs, this fra in a finite gamemework can be used to evolve strategies that outperform other strategies in the game. In one recent example, Harper et al. used evolutionary algorithms to evolve automata against a library of known strategies (e.g. Fool Me Once, LookerUp, DFA), resulting in trained strategies with cooperative tendencies \cite{harper2017reinforcement}. 

We are interested in examining the emergence of cooperation in the context of co-evolution, where rather than evolving automata with respect to a specified objective or pre-determined adversaries, we allow the candidate automata to play games of IPD with co-evolving cohorts and make selection based on their relative fitness. This characterization of cooperation --- not necessarily as an optimal strategy, but as an emergent phenomenon borne out of evolutionary and population pressure ---  was laid out by Axelrod in his seminal work \emph{Evolution of Cooperation}  \cite{axelrod_evolution_1984}. Since then, genetic algorithms (GA) have been predominantly used in literature to evolve IPD-playing agents, such as finite-state automata \cite{MILLER199687}, bounded-state automata \cite{marks_breeding_1992}, and string-based representation of iterated strategies \cite{doi:10.1002/cplx.1030}. 

We extend upon previous works by introducing probabilistic automaton (PA), specifically the \emph{stochastic Moore machine} into the IPD evolutionary framework. In doing so, we proposed a novel mutation mechanism that allows real-value probabilities and their correspondent behaviors to be evolved. 
This modification enables the representation of a behaviorally richer class of automata in the study of cooperative strategies. For instance, a finite-state automaton behaves with bounded rationality in which the number of states equals the recall length, but a probabilistic automaton can have unbounded lengths of recall based upon stochasticity, thus allowing an agent to make its decision based upon an indefinite length of past interactions. A deterministic automaton has also been proven to be reducible to a probabilistic automata, making probabilistic automata a strictly more general class \cite{rabin_probabilistic_1963}. Recent developments in stochastic strategies led to the discovery of \emph{zero-determinant strategies}, a form of `memory-one' stochastic strategy that can unilaterally determine the payoff of their opponents \cite{press_iterated_2012}. However, zero-determinant strategies have been shown to be evolutionarily unstable, and are thus not pertinent to our discussion \cite{adami_evolutionary_2013}. 

The models and experimental schema introduced in this paper build upon the idea that cooperation is an emergent phenomenon arising from the interplay of evolution, incentive structure and repeated interactions, and that the modification of each would have significant effects on the emergence and persistence of cooperation. We first develop and justify representative models for each of these three components, and then run computer simulations to observe and study the emergence and stability of cooperation. 

The rest of the paper is organized as follows. Section II delineates our modeling of sequential behavior and evolutionary mechanisms. Section III elaborates upon the evolutionary mechanisms in the context of evolutionary theories. Section IV discusses the results obtained from our evolutionary simulation, and Section V concludes the paper with proposals for follow-up studies and future works. Section VI (Appendix) includes graphs and illustrations pertinent to the discussions in the rest of the paper.

\section{Modeling} \label{modeling}
\subsection{Sequential Behavior}
Our proposed model uses \emph{stochastic Moore machines} to represent sequential behaviors. SMMs have been studied in the context of strategy modeling in literature, and in fact there exist algorithms such as MDI \cite{thollard_probabilistic_2000}, GIATI \cite{casacuberta_machine_2004} that can be used to approximate and predict the stochastic automaton used by one's opponent \cite{cebrian_learning_2011}. A stochastic Moore machine is a tuple $M=\langle S, s_0, \Phi, \Lambda, T, G \rangle$ where:
\begin{itemize}
    \item $S=\{s_0, ..., s_n\}$ is a finite set of states.
    \item $s_0$ is the initial state.
    \item $\Phi$ is the set of input alphabets (observations).
    \item $\Lambda$ is the set of output alphabets (actions).
    \item $T : S \bigtimes \Phi \mapsto [0,1]^{n+1}$ is the transition probabilities between states.
    \item $G : S \mapsto \Lambda$ is the mapping between state and action.
\end{itemize}

This automaton, when at state $s_i \in S$, takes in input $\phi \in \Phi$ and moves to any state $s_j$ with probability $p_j(s_i,\phi)$, and performs action $\lambda \in \Lambda$. For this paper, $\Phi=\{C,D\}$ and $\Lambda=\{C,D\}$ represent each agent's choice between cooperation (C) and defection (D).

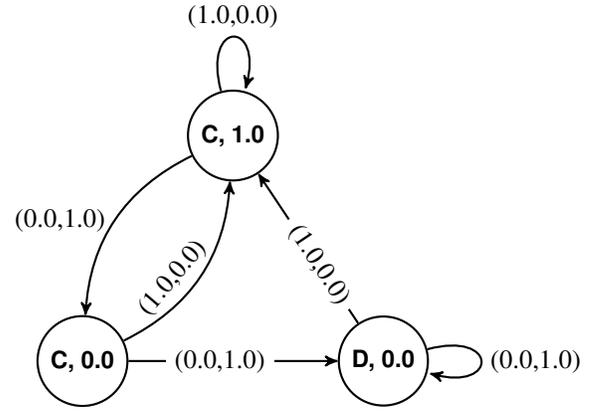
\begin{figure}
\centering
\begin{tikzpicture}[->,>= stealth', thick, auto, main node/.style={circle,draw,font=\sffamily\small\bfseries}]
\node[main node] (1) {C, 1.0}; 
\node[main node] (2) at (-2,-3) {C, 0.0};
\node[main node] (3) at (2, -3) {D, 0.0};

\path[]
    (1) edge [bend right] node[left] {(0.0,1.0)} (2)
        edge [loop above] node {(1.0,0.0)} (1)
    (2) edge [bend right, sloped, fill=white, pos=0.4] node {(1.0,0.0)} (1)
        edge node[left, pos=0.7, fill=white] {(0.0,1.0)} (3)
    (3) 
        edge node [right, sloped, fill=white, pos=0.7] {(1.0,0.0)} (1)
        edge [loop right] node {(0.0,1.0)} (3);

\end{tikzpicture}
\caption{A stochastic Moore machine representing two-tits-for-a-tat. The pair of numbers (A,B) on an edge represents the likelihood of transiting to the destination state given that the opponent \emph{cooperated} or \emph{defected} respectively. The number in the node indicates the probability of starting at that node.}
\end{figure}
In order for the probabilistic automaton (PA) to be valid, the transition probabilities must satisfy the following condition:
\begin{equation}
    \sum_j p_j(s_i, \phi) = 1 \quad \forall \phi
\end{equation}
and 
\begin{equation}
    p_j(s_i, \phi) \geq 0 \quad \forall \phi, j
\end{equation}

The number of states in the PA thus represents the size of memory of the agent, and a larger number of states would allow for the description of richer behavioral schemas. Ben-Porath \cite{ben-porath_repeated_1993} utilized the minimal number of states necessary to define a strategy as its complexity measure, hence we can also treat the number of states as a measure of strategy complexity. 

Our choice of SMMs to represent sequential strategies was inspired by previous works on automata \cite{fogel_artificial_1966} \cite{ben-porath_repeated_1993}. While most work on this topic has used deterministic automata, we believe that a stochastic model is necessary in solving certain games. For instance, it allows for fair coordination in games with dominance structure, as will be discussed in Section IV (c). 
\subsection{Evolutionary Operators}
The stochastic Moore machine modeling of sequential behavior is ideal for use in evolutionary algorithms, as it readily gives a genotypic representation of behavior. Specifically, we can represent the evolvable genotype of a Moore machine as $\langle T, \Theta, G \rangle$, where:

\begin{itemize}
    \item $T : S \bigtimes \Phi \mapsto [0,1]^{n+1}$ is the transition probabilities between each pair of states.
    \item $\Theta=\{\theta_0, ..., \theta_n\}$ is the probability vector of starting at each state, such that $p(s_0=s_i)=\theta_i$.
    \item $G : S \mapsto \Lambda$ is the mapping between each state $s_i$ and its corresponding action
\end{itemize}

Since both $T$ and $\Theta$ are comprised of real numbers, we can develop various numerical functions to manipulate these quantities and model certain phenomena. For instance, we model mutation as the addition of Gaussian noise upon $T$ and $\Theta$, which implicitly changes the behavior of the agent by modifying the transition and initial probabilities of its states. 

\begin{figure}
    \centering
    \includegraphics[width=0.5\textwidth]{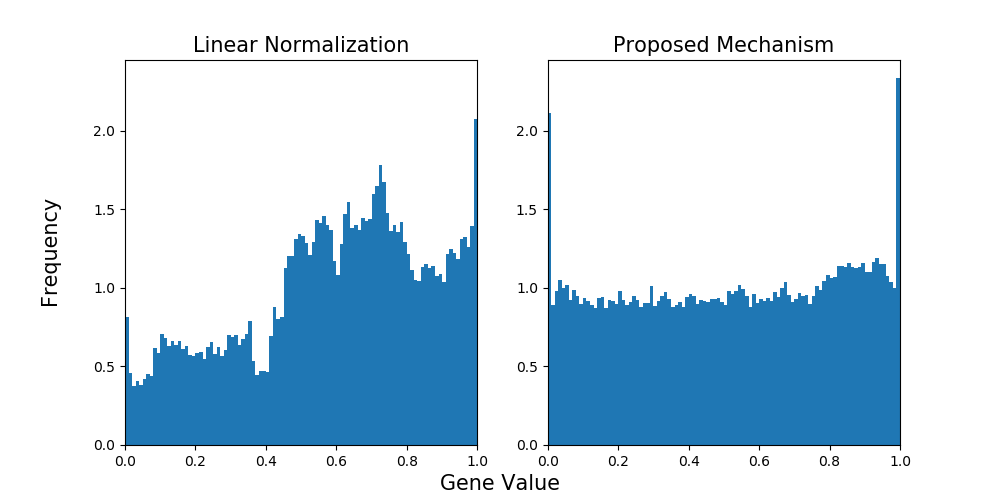}
    \caption{The sampling distribution of values of a single gene in a 2D vector across 100,000 iterations under linear normalization mutation (left) and our proposed mutation mechanism (right).}    
    \label{fig:single_run_value_dist}
\end{figure}

The difficulty in applying Gaussian addition onto probability vectors is that it may break the constraints defined in Equation 1 and 2. In past works, this was resolved by normalizing the probability vector after additive mutation \cite{harper2017reinforcement}. However, this approach leads to asymmetry in the direction of mutation, because a negative mutation ($\delta < 0$) is scaled up by normalization while a positive mutation ($\delta > 0$) is scaled down. Figure \ref{fig:single_run_value_dist} illustrates the net result of this behavior. The presence of such a bias makes the mutation mechanism unsuitable for evolutionary simulation due to its compounding effects on the population genome.

To mitigate this effect, and to improve upon previous methods we propose a novel mutation mechanism for probability vectors. We know that in order for transition probabilities to sum up to one, net changes to the probabilities vector must be zero, i.e. $\sum \mathcal{N}(0, \sigma^2)=0$. We achieve this by pairing randomly-chosen entries of the probability vector $p(s,\phi)=T_{s\phi}$, where $s \in S, \phi \in \Phi$, when applying Gaussian mutation. That is, when performing a mutation operation, we pick $i,j$ such that $i \neq j$ to perform $mutate(i,j,s, \phi)$, defined as 
\begin{gather}
    \begin{aligned}
    \delta &= \mathcal{N}(0, \sigma^2)  \\
    p_i(s, \phi) &= p_i(s, \phi) + \delta \\
    p_j(s, \phi) &= p_j(s, \phi) - \delta
    \end{aligned}
\end{gather}

This operation can be performed indefinitely with the probabilities sum remaining invariant. However, the non-negative constraint on probability values as expressed in Equation 2 may still be violated when the magnitude of a negative mutation is larger than the current probability value. To mitigate this, we make adjustments to the value of $\delta$ in $mutate(i,j,s, \phi)$ prior to applying the mutation, 
\begin{equation}
    \delta = min(\delta, 1-p_i(s, \phi),p_j(s, \phi))
    \label{eq:clipping}
\end{equation}

\begin{figure}[!ht]
    \centering
    \includegraphics[width=0.5\textwidth]{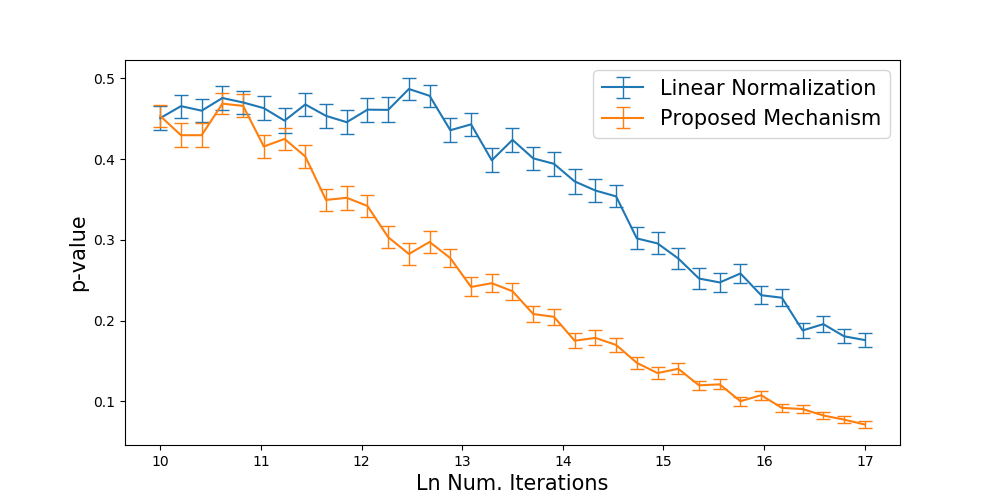}
    \caption{The p-value of the highest-valued distribution fitted against uniform distribution, along log number of iterations. Lower values imply better fit, with 0.0 representing uniform distribution.}    
    \label{fig:value_dist_p_value}
\end{figure}

When applied to a 2D vector, our proposed mechanism leads to a near-even distribution of values for a gene, as illustrated in Figure \ref{fig:single_run_value_dist} (b). It is also empirically shown to be generalizable to N-dimensional vectors (refer to the Appendix). On the basis that an unbiased mutation mechanism should not favor any particular gene, we also devised a metric to compare the performance of our mechanism with linear normalization by measuring the closeness of fit between the observed frequency of genes with the highest value under mutation operations (hereon referred to as highest-value frequency) with that of a uniform distribution. We defined the null hypothesis as the highest-value frequency being drawn from a non-uniform distribution, and calculated its p-value as one minus the probability obtained through Chi-square test. Hence, a low p-value indicates a greater confidence in an unbiased distribution. Fig. \ref{fig:value_dist_p_value} shows that our proposed mechanism outperforms the conventionally-used mutation mechanism in this regard, maintaining low p-values across iterations. 

We acknowledge that this operation may still introduce bias at the boundaries of the interval $[0,1]$ due to the clipping effects of Equation \eqref{eq:clipping}, and hope to address this issue in future works.
\subsection{Modeling Interactions}
In order to model interaction we look at interactions between agents from an evolutionary game theory perspective, where each agent seeks to maximize its utility to increase its likelihood of survival and propagation. Although we do not explicitly program this objective into the agents, it emerges out of evolutionary pressure, as agents that perform well in their interactions with other agents are assigned higher fitness scores and thus can potentially stick around longer in the simulation.

We model the environment as a fixed-size population $\{M_0, ..., M_{n-1}\}$, where each $M_i$ represents one of the $n$ agents. To compute the fitness value of an agent, we pit it against every other agent $M_j$ in the population in an iterated game. We use a \emph{relative fitness} metric based on each agent's interactions with the rest of the population. The scores from a pairwise interaction can be calculated through repeated simulation. In every interaction, each pair of agents will play an iterated 2x2 normal form game for \emph{k} iterations, and in each iteration the states $s_{it}$ and $s_{jt}$ of both SMMs will be tracked and updated. The payoff $u_i$ for a certain agent $M_i$ is then calculated as follows, where $M_j$ is the opponent:
\begin{equation}
    u_i(M_i, M_j) = \sum_{t=0}^{k-1}{u_i(G_i(s_{it}), G_j(s_{jt}))}
    \label{eq:payoff}
\end{equation}

Since the approach described above is computationally expensive, we devised a set of matrix operations that yield the time-average payoff of (\ref{eq:payoff}) at equilibrium. The implementation details of which are given in the Appendix.

We compute the cumulative score of an agent by adding up its pairwise score against all other agents,
\begin{equation}
    C(M_i) = \sum_{j\neq i}u_i(M_i,M_j)
\end{equation}
which we then use as a metric of fitness in determining their survival and reproductive chances.

The chosen 2x2 normal form games are based on their game-theoretic qualities. We decided to investigate Prisoner's Dilemma because of its interesting properties and for its lasting influence on the game theory community. We also included results from Chicken, Stag Hunt, and Battle in the Appendix.

\section{Evolutionary Mechanisms}
The application of non-goal-directed and Darwinian evolution in simulating strategy evolution is limited in the literature, and thus is taken as the prime focus of this paper. We investigated whether in the absence of heuristics, cooperation can emerge organically. In this section, we discuss evolutionary paradigm used to simulate evolution. 

Selection mechanisms can be introduced into the simulation in two dimensions --- survival selection and reproductive selection. Both sources are evidently present in the natural world; organisms invest resources to ensure their survival and to find a mate. Peacocks exemplify the trade-offs that organisms make to adapt to these two pressures. The cumbersome feather display of the male peacock hurts its chances of outrunning a predator but increase its mating chances. More often though, the reproductive fitness of an organism is directly tied to its survival fitness, hence we used the same fitness metric in determining an agent's reproductive and survival chances. It is a simplified model, but nonetheless serves to build a platform on which more accurate representations can be added. 

Survival and reproductive selection are used under a different context in the field of evolutionary algorithms, a discipline distinct from biological evolutionary studies. In the field of evolutionary algorithms, selection mechanism is seen as a heuristic that guides a stochastic search down a non-convex surface, where too much pressure can lead to sub-optimal convergence and too little can result in failure to converge. De Jong gives a summary of this trade-off in \cite{de_jong_evolutionary_2016}.

We acknowledge both approaches to modeling evolutionary pressure, one in terms of representing the natural world and the other as an optimization strategy. For this paper, we are interested in investigating whether cooperative tendencies emerge in nature by necessity or by chance. We investigated if the emergence of cooperation is \emph{inevitable} given a particular set of environmental variables. Hence, when developing our model we kept both perspectives in mind. We hoped to gain insights into the natural origin of cooperation by studying the efficacy of cooperative tendencies in surviving artificial and simplified models of evolution. 

As discussed above, the two knobs that can be tweaked are survival and reproductive selection mechanisms. Strict selection mechanisms can be used as exploitative strategies, because offsprings of high fitness agents can be greedily chosen to drive up the average fitness of a population. On the other hand, loose selection mechanisms, such as the introduction of stochasticity or randomness in selection, can be used to encourage exploration and prevent sub-optimal convergence. The standard approach in the field of evolutionary algorithms usually involves balancing these two strategies by using strict mechanisms in reproductive selection and stochastic mechanisms in survival selection or vice versa. 

To be comprehensive, we considered various combinations of reproductive and survival selection methods, listed in Table \ref{table:selection_mechanisms}. We also experimented with letting the parents compete with their own offsprings for survival, designated here as \emph{overlap}. 

\begin{table}[h]
\begin{center}
\begin{tabular}{c c c c c c}
Index & Reproductive Sel. & Survival Sel. & Overlap\\
\hline 
(a) & truncation & truncation & yes\\
(b) & roulette & truncation & yes\\
(c) & truncation & truncation & no\\
(d) & uniform & truncation & no\\
(e) & truncation & uniform & yes\\
(f) & roulette & uniform & yes\\
(g) & truncation & uniform & no\\
(h) & roulette & uniform & no\\
(i) & uniform & uniform & yes\\
\end{tabular}
\end{center}

\noindent\textbf{truncation}:\tabto{5em} Select N agents with the highest fitness scores.

\noindent\textbf{roulette}:\tabto{5em} Select (without replacement) N agents with probability proportional to their fitness scores.

\noindent\textbf{uniform}:\tabto{5em} Select (without replacement) N agents randomly.

\caption{Selection Mechanisms}
\label{table:selection_mechanisms}
\end{table}

\section{Experiments}

\begin{figure*}
        \centering
        \begin{subfigure}[b]{1\textwidth}
            \centering
            \includegraphics[width=\textwidth]{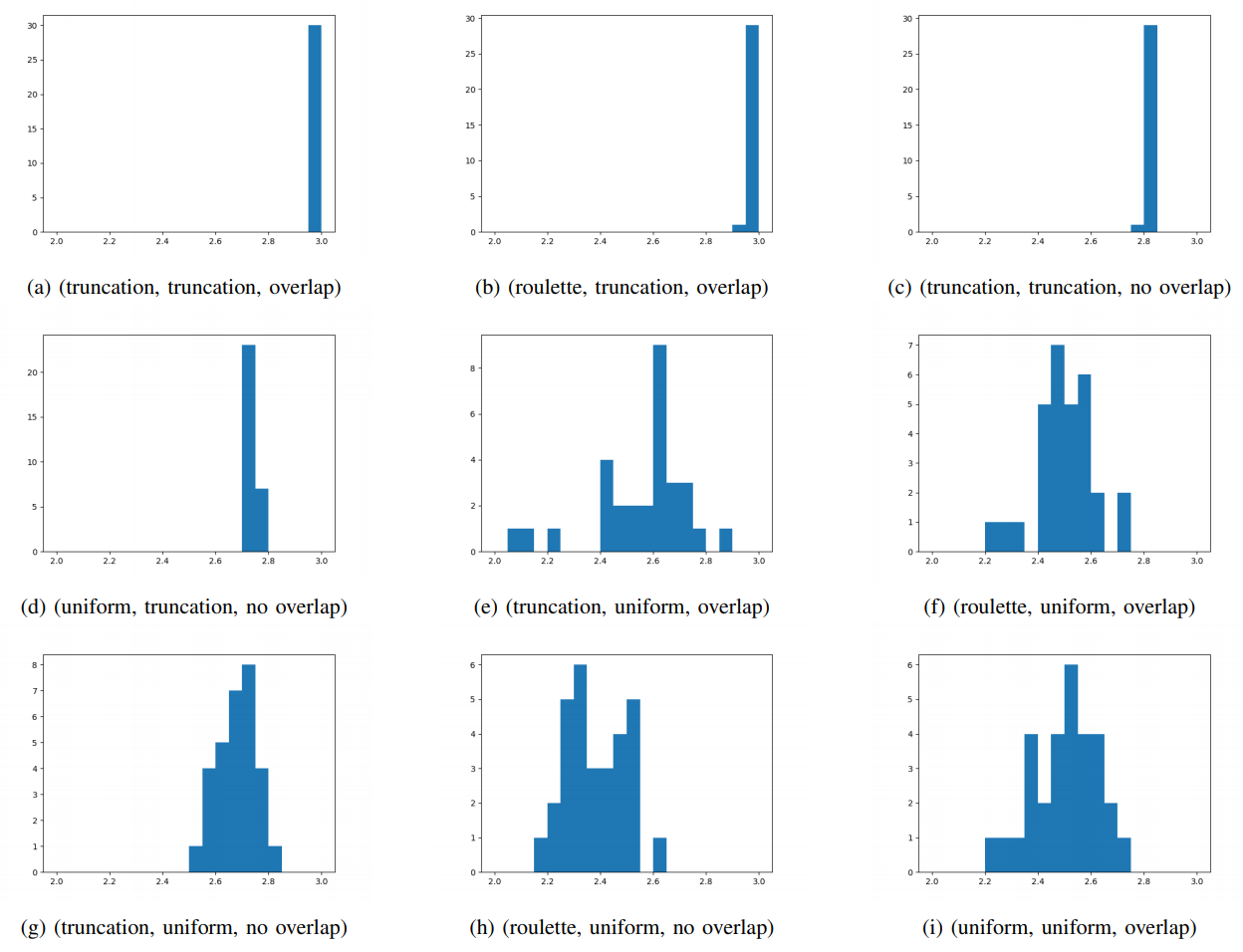}
            \label{fig:mean and std of net14}
        \end{subfigure}
        \caption[]
        {\small Distribution of average agent score in 30 trials across different combination of selection mechanisms. It is evident that the use of truncation method to select for survival greatly increases the likelihood that the resulting population converges to  cooperative strategies, with mean scores nearing 3. \emph{x}: mean agent score, \emph{y}: number of trials, \emph{label}: (parent selection, survival selection, overlap)} 
        \label{fig:mean and std of nets}
    \end{figure*}

For each selection paradigm, we ran 30 simulations. In each simulation, we tracked the interactions of 20 different agents across 1000 iterations, simulating $20 \choose 2$ pairwise interactions per iteration. We also recorded the mean score for each simulation, which was calculated as the aggregate score obtained from plays of IPD, averaged across agents in the population and across iterations. The first 200 iterations were discarded to allow a grace period for agents to converge on distinct strategies from random initial configurations. To analyze the emergence of cooperation, we plotted a histogram of the mean scores across all simulations corresponding to each set of selection mechanisms, as shown in the Figure \ref{fig:mean and std of nets}. 

\begin{table}[!ht]
    \centering
    
    \begin{tabular}{|c|c|c|}
        \hline
         & D & C  \\
         \hline
        C & 1,4 & 3,3 \\ 
        \hline
        D & \bf{2,2} & 4,1 \\
        \hline
    \end{tabular}
    \caption{Payoff Matrix for Prisoner's Dilemma. Each cell corresponds to the pair of utility gained $u(G_i, G_j)$ by player $i$ and $j$ for each combination of players actions}
    \label{table:PD_payoff}
\end{table}
The mean score is correlated with the prevalence of cooperative strategies as follows: referring to the payoff matrix in Table \ref{table:PD_payoff}, we see that pairwise mean score is lowest when both defect ($2.0$), middling when one cooperates and the other defects ($2.5$), and highest when both cooperate ($3.0$). From averaging the pairwise mean scores, we can thus place a bound on the proportion of interaction-types that have occurred by appealing to some simple math. A mean score of $3.0$ implies that every agent interaction resulted in mutual cooperation, $2.75$ implies that at least $50\%$ of the interactions were mutual cooperation, and $2.5$ or lower implies that mutual cooperation was a minority of all the interactions.

We found that cooperation consistently emerged in all instances where strong survival pressure was present (Figure \ref{fig:mean and std of nets}(a)-(d)), and failed to dominate in the absence of it (Figure \ref{fig:mean and std of nets}(e)-(h)). These experiments showed that cooperation can spontaneously emerge under specific conditions.

We also included a control group (Figure \ref{fig:mean and std of nets}(i)) with randomized parent and survival selection to validate the mutation mechanism. The resulting mean scores form a Gaussian around $2.5$, which corroborates that there is no systemic bias. 

\subsection{Analysis of Individual Simulation}
To better understand the conditions leading to the emergence of cooperation, we analyzed the evolutionary graph of one particular selection paradigm with strong survival pressure (Paradigm (a) in Table \ref{table:selection_mechanisms}).

\begin{figure}[h]
\centering
\includegraphics[trim =0mm 0mm 0mm 0mm, clip,scale=0.5]{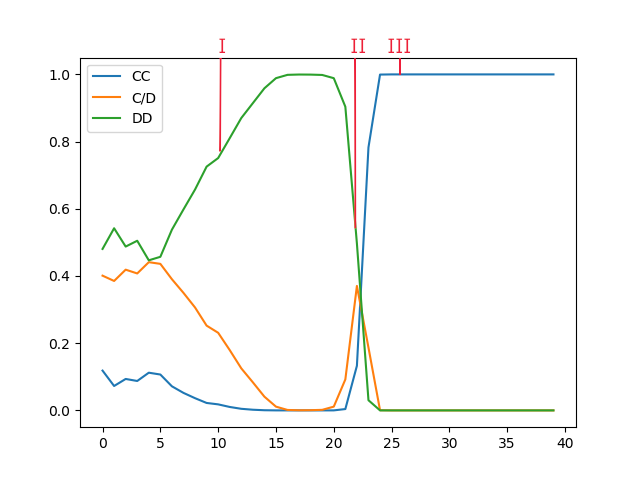}
\caption{\small This graph shows the ratio of interactions for each type (CC, CD, or DD) averaged across a window of 5 generations for smoothing, for a total of 200 generations. We see the emergence of exploitative strategy at Phase I, exploitative strategy at Phase II and trigger strategy at Phase III.}

\label{fig:single_run}
\end{figure}

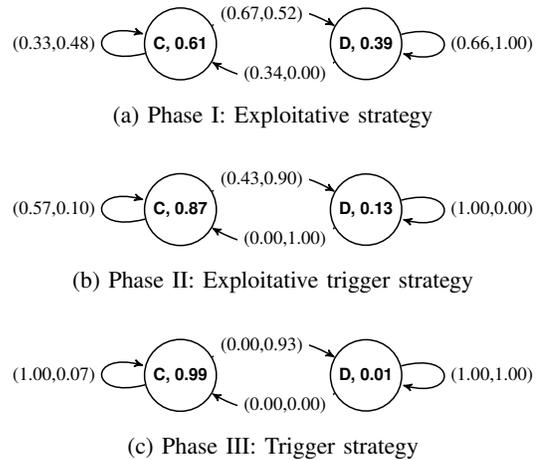
\begin{figure}[ht]
\centering
   \begin{subfigure}[b]{0.4\textwidth}
       \resizebox{\linewidth}{!}{
       \begin{tikzpicture} [auto, node distance=3cm, ->, >=stealth', thick, main node/.style={circle,draw,font=\sffamily\small\bfseries}]
        \node[main node] (1) {C, 0.61};
        \node[main node] (2) at (3.5,0.0) {D, 0.39};
        
        \path[] 
        (1) edge[bend left] node[left, pos=0.8, fill=white] {(0.67,0.52)} (2)
        edge [loop left] node {(0.33,0.48)} (1)
        (2) edge[bend left] node[right, pos=0.8, fill=white] {(0.34,0.00)} (1)
        edge [loop right] node {(0.66,1.00)} (2);
        \end{tikzpicture}
        }
        \caption{Phase I: Exploitative strategy}
   \end{subfigure}
    \\[3ex]
   \begin{subfigure}[b]{0.4\textwidth}
       \resizebox{\linewidth}{!}{
       \begin{tikzpicture} [auto, node distance=3cm, ->, >=stealth', thick, main node/.style={circle,draw,font=\sffamily\small\bfseries}]
        \node[main node] (1) {C, 0.87};
        \node[main node] (2) at (3.5,0.0) {D, 0.13};
        
        \path[] 
        (1) edge[bend left] node[left, pos=0.8, fill=white] {(0.43,0.90)} (2)
        edge [loop left] node {(0.57,0.10)} (1)
        (2) edge[bend left] node[right, pos=0.8, fill=white] {(0.00,1.00)} (1)
        edge [loop right] node {(1.00,0.00)} (2);
        \end{tikzpicture}
        }
        \caption{Phase II: Exploitative trigger strategy}
   \end{subfigure}
    \\[3ex]
   \begin{subfigure}[b]{0.4\textwidth}
       \resizebox{\linewidth}{!}{
       \begin{tikzpicture} [auto, node distance=3cm, ->, >=stealth', thick, main node/.style={circle,draw,font=\sffamily\small\bfseries}]
        \node[main node] (1) {C, 0.99};
        \node[main node] (2) at (3.5,0.0) {D, 0.01};
        
        \path[] 
        (1) edge[bend left] node[left, pos=0.8, fill=white] {(0.00,0.93)} (2)
        edge [loop left] node {(1.00,0.07)} (1)
        (2) edge[bend left] node[right, pos=0.8, fill=white] {(0.00,0.00)} (1)
        edge [loop right] node {(1.00,1.00)} (2);
        \end{tikzpicture}
        }
        \caption{Phase III: Trigger strategy}
   \end{subfigure}
    \\[3ex]

    \caption{A sample SMM configuration for each labeled window of generations. The pair of numbers (A,B) represents the likelihood of transiting to the destination state given that the opponent cooperated or defected respectively. }
    \label{fig:SMM}
\end{figure}

The evolutionary graph of one simulation instance is illustrated in Figure \ref{fig:single_run}. Here, we noticed three distinct phases that the population went through, indicated by the red lines in the figure. By comparing the SMMs present at the windows marked by the respective phase lines (each window is defined as an average of 5 iterations), we built the following evolutionary narrative:

In the first phase (Figure \ref{fig:SMM} (a)), exploitative strategies prevailed in which cooperation was mixed with occasional defection. To counter these exploitative strategies, the agents became increasingly unforgiving towards defection, resulting in a steady increase in defection rates from window 5 to window 13. From window 13 onwards, all the SMMs had a near-zero chance of returning to the cooperative state after they had seen one defection (ie. become triggered), as can be seen in Figure \ref{fig:SMM} (b). As a result, exploitative strategies became undesirable, and the rate of defection following mutual cooperation began to dwindle. This eventually lead to more consistent cooperation, and the perfecting of the grim trigger strategy at around window 22 (Figure \ref{fig:SMM} (c)).  

Grim strategy, a member of the larger class of trigger strategies, is an example of an automaton with trapping states in which a past action is remembered for the rest of the interaction \cite{miller_evolution_1988}. We observed that this convergence towards the grim trigger strategy occurred consistently across most simulations in which cooperation emerged. When interacting with other strategies, such as pure Tit-for-Tat or occasionally-exploitative Tit-for-Tat, grim strategy punishes any defection with unforgiving retaliation but rewards cooperation with reciprocation. It is thus in any other strategy's best interest to not defect against the grim strategy. 

Moreover, we observed that after the emergence of grim strategy, the population remained stable and undisturbed by mutant invaders. The transition probabilities in the post-cooperation generations reveal that mutations do occur in the form of increasing forgiveness or exploitation rate, but these deviant strands did not prove to be beneficial to the agent and are therefore selected away before they can effectively change the population makeup. The population converged at an \emph{evolutionarily  stable state} as was defined by Smith: a population in which the genetic composition is restored by selection after a disturbance, provided the disturbance is not too large \cite{smith_evolution_1982}. This however does not preclude the possibility that the existing strategies could be invaded by genetically distinct strategies, since every mutant in the evolutionary paradigm shared genetic ties with their parents. Thus, we could not conclusively verify the stronger assertion that the population converged at an \emph{evolutionarily stable strategy}. 

Another aspect that we investigated was the polymorphism of the evolved population, that is, whether we had not just a single strategy but rather a composition of strategies resisting disturbance. The existence of such heterogeneous populations is discussed in the next subsection.

\subsection{Diversity of Co-existing Strategies}
We considered the possibility that the population may consist of distinct strategies that work well in the presence of each other -- a group of \emph{mutually compatible} strategies in an evolutionarily stable state. We therefore propose two approaches to the categorization of behaviorally isomorphic agents.

1. Clustering: If we were to define a function that measures the similarity in behavior between two agents, which can be thought of as a Kernel function, we would be able to make use of the wide gamut of clustering algorithms for our goal of clustering agents according to their behavior. Unfortunately, computing the similarity or showing equivalence of two finite-state automata has been shown to be undecidable \cite{griffiths_unsolvability_1968}. Therefore, the best we could do is to interpret our agents as finite-state machine graphs and use graph similarity metrics, of which there are many \cite{gallagher_matching_2006}, to compute the similarity between two agents. We plan to provide an analysis of behavioral profiles in future studies. 

2. Behavioral Isomorphism: Since measuring the structural similarity between two non-deterministic machines has been shown to be undecidable, we looked at measuring behavioral similarity instead. In our experiments, we observed that the actions played by the agents in each interaction increased greatly in homogeneity after the first few dozens of iterations as agents that took different and more inferior actions were quickly selected away. Our observation is consistent with the concept of convergence as defined by Creedy \cite{creedy1992recent}, who stated that convergence is reached when the different machines in the population play the same actions for the duration of all repeated games. These machines can be therefore can be treated as behaviorally (not necessarily structurally) identical.

Building on this definition, we propose a simple metric to quantitatively measure the behavioral homogeneity of a population at $t$-th iteration by considering the equilibria of pairwise interactions: 

\begin{equation}
    D(t) = \dfrac{\sum_{i,j, i\neq j}\Vert a_{i} - a_{j}\Vert_2^2}{n(n-1)}
\end{equation}
where $a_i$ is the action probability of $M_i$ at horizon, and $n$ is the number of agents. 

This equation finds the average mean-squared error of the horizon action probability between each pair of agents. If in horizon the pair of interacting agents have the same probability of cooperating/defecting, then they are said to be behaviorally similar under this metric. Some information may be lost due to this generalization, for instance we would not be able to differentiate a grim strategy from a Tit-for-Tat because Tit-for-Tat will be able to emulate the behavior of grim strategy within one move. But in the context of finding convergent behavior this metric still serves to inform us whether certain agents behaviorally dominate others. 

To test the merit of our proposed metric, we computed it across simulations of 200 generations. We observed that $D(t)$ dropped to near zero-value in around 10 generations on average, implying that the different machines began to act homogeneously early on, and that we have a behaviorally isomorphic population by the end of the simulation. Therefore, a single agent can be considered an accurate representation of the population when discussing our results. 

\subsection{Other Symmetric Games}
To demonstrate the generalizability of our proposed mutation mechanism and experimental setup, we ran simulations for other well-known games such as Chicken, Stag Hunt, and Battle under selection paradigm (a). We have included the resulting simulation graphs as well as one SMM agent (sampled from near-homogeneous populations) for each game in the Appendix. 

\begin{figure*}[t] 
\includegraphics[height=155.0pt, width=1.0\textwidth]{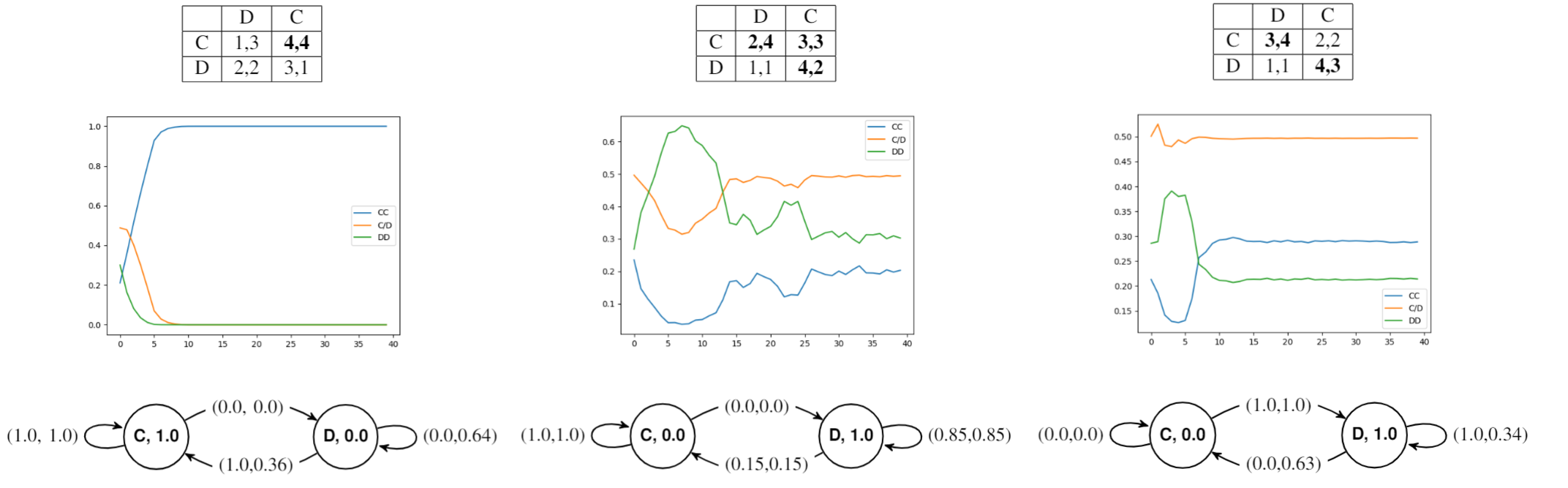}
\caption{(left to right) Stag Hunt, Chicken, Battle}
\label{fig:other_games}
\end{figure*}

We chose these games because they are symmetric and have interesting Nash equilibria. For instance, the Battle game (Figure \ref{fig:other_games}) presents a unique coordination challenge for its players, as the Pareto outcome arises when one player takes the opposite action of the other player (e.g. CD or DC), with the cooperator receiving the smaller payoff. 
It is thus necessary for any surviving strategy to strike a balance between compromising and asserting dominance in the game.

The SMMs that evolved from our experimental setup achieved this balance through effective use of randomization. For instance, when DD occurs, the agents have a 63\% chance of switching to cooperation in the next round. The likelihood that they both end up with an undesirable outcome (CC or DD) is thus $0.63 * 0.63 + 0.37 * 0.37 \approx 0.53$, allowing them to achieve Pareto outcomes around 50\% of the time, equivalent in performance to that of the mixed strategy Nash equilibrium. 

Similarly, the evolved SMMs of Chicken and Stag Hunt also have an average performance equal to their respective mixed strategy Nash equilibria. These examples thus reveal that the proposed approach can be used to discover equilibrium strategies in games beyond IPD, and do so efficiently in very few iterations.



\section{Conclusion}

In this paper we proposed a novel mutation mechanism and demonstrated that it significantly improves over existing methods for evolving probability vectors of arbitrary dimensions. We used this mechanism in conjunction with our simulation setup and studied the emergence of cooperation. We can conclude that cooperation can robustly emerge and persist even under the threat of selfish behaviors. The shaping of the cooperative and defecting aspects of trigger strategies, motivated by avoiding retaliation and countering exploitation respectively, shows that cooperative behaviors can be deemed as rational adaptations when viewed from an evolutionary perspective. We also tested our approach on different games and concluded that our methods are indeed generalizable and in fact can be used to find strategies equivalent to mixed strategy Nash equilibria. 


In future works, we aim to provide a rigorous mathematical backing on the universality and limitations of the proposed mutation mechanism. We also intend to explore the use of our methodology to efficiently analyze game-theoretic properties such as mixed strategy Nash Equilibrium of iterated games. 

The code and implementations of our mutation mechanism, simulation setup and homogeneity metric can be found at: \texttt{https://github.com/jinhongkuan/evol-sim}

\section{Appendix}

\subsection{Pairwise Gaussian Mutation}

\begin{figure}[!h]
    \begin{subfigure}[b]{1.0\textwidth}
    \includegraphics[width=0.5\textwidth]{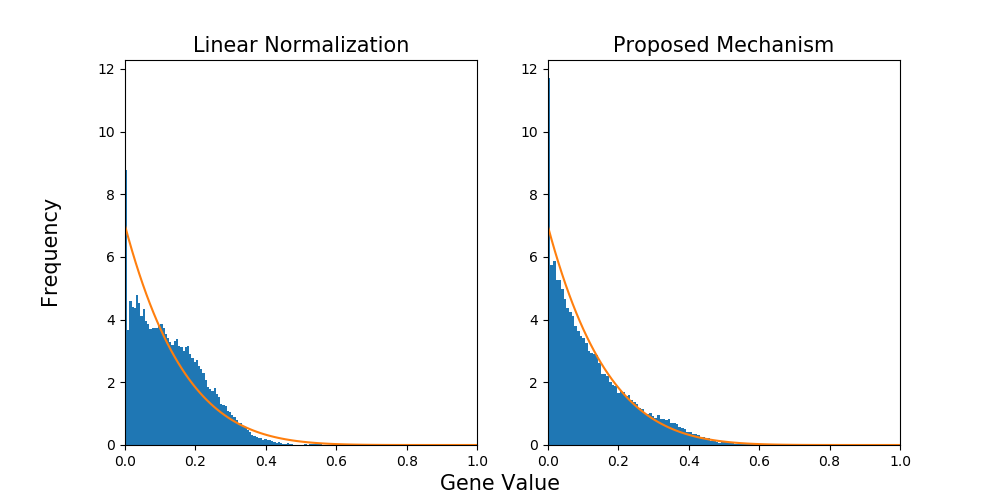}
    \end{subfigure}    
    \caption{The sampling distribution of values of a single gene in 8D vectors across 100,000 iterations. The orange curve represents the ideal distribution with no bias.}    
    \label{fig:value_dist}
\end{figure}

We assessed the proposed mechanism by comparing fit of its resulting gene-value distribution with the theoretical distribution of uniformly-sampled probability vectors (derivations can be found in the GitHub repository mentioned above), 
\begin{align}
    P(p_i = k) = (\vert \mathbf{p} \vert -1)(1-k)^{\vert \mathbf{p} \vert -2} \label{eq:marg_density}
\end{align}
Figure \ref{fig:value_dist} illustrates the simulation result for 8D probability vectors under both linear normalization and the proposed mechanism. As indicated by \eqref{eq:marg_density}, the theoretical distribution is in fact not uniform in N-dimensional spaces where $N>2$, but instead forms a polynomial curve. 

The proposed mechanism yielded a good fit to the theoretical distribution, achieving $p<0.05$ fit across all tested dimensions (2D to 8D) at 100,000 iterations. Hence, we assert that the applicability of the mechanism generalizes to higher dimensions. 

\subsection{Matrix Operations for IPD payoff}
Let $\mathbf{p}_{t}$ represent the state probability vector of a SMM at a given time $t$, such that $\mathbf{p}_0=\Theta$ which is the probability vector of starting at each state. Referring to the definition of an evolvable genotype of Moore machine $\langle T, \Theta, G \rangle$, we model the interaction between two SMM as follows: 
\begin{gather}
\begin{aligned}
    f(T,\mathbf{a}_{jt}) &= \sum_{k=0}^{|\Phi|-1} T(\phi_k)a_{jtk} \\
    \mathbf{p}_{t+1} &= f(T,\mathbf{a}_{jt})^T \mathbf{p}_t\\
    \mathbf{a}_{i(t+1)} &=  G (\mathbf{p}_t)
\end{aligned}
\end{gather}
where $T(\phi_k)$ is a matrix of state-to-state transition probabilities given input $\phi_k$, and $\mathbf{a}_{it}$ is the action probability vector of machine $M_i$ at iteration $t$. The same applies for machine $M_j$, with $i$ and $j$ flipped in the above equation. Empirically, we found that the above matrix operations converges in around 5 operations, and the resulting probability distribution gives accurate horizon values as tested by running repeated simulations for thousands of repetitions. We used this method to run the simulations discussed in this paper.

\bibliographystyle{IEEEtran}
\bibliography{ESS.bib}

\end{document}